\documentclass[showpacs,amssymb,twocolumn,floatfix,aps,notitlepage]{revtex4-1}

\usepackage{amssymb,amsfonts,amsmath,bm}
\usepackage{graphics,graphicx}
 
\begin{document}

\title{Poisson-Boltzmann thermodynamics of counter-ions confined by 
curved hard walls}

\author{Ladislav \v{S}amaj}
\affiliation{Institute of Physics, Slovak Academy of Sciences, Bratislava, 
Slovakia}
\author{Emmanuel Trizac}
\affiliation{LPTMS, CNRS, Univ. Paris-Sud, Universit\'e Paris-Saclay, 91405 Orsay, France}

\begin{abstract}
We consider a set of identical mobile point-like charges (counter-ions) 
confined to a domain with curved hard walls carrying a uniform fixed 
surface charge density, the system as a whole being electroneutral.
Three domain geometries are considered: a pair of parallel plates, 
the cylinder and the sphere.
The particle system in thermal equilibrium is assumed to be described by 
the nonlinear Poisson-Boltzmann theory. 
While the effectively 1D plates and the 2D cylinder have already been solved,
the 3D sphere problem is not integrable.
It is shown that the contact density of particles at the charged surface
is determined by a first-order Abel differential equation of the second kind 
which is a counterpart of Enig's equation in the critical theory 
of gravitation and combustion/explosion. 
This equation enables us to construct the exact series solutions of 
the contact density in the regions of small and large surface charge densities.
The formalism provides, within the mean-field Poisson-Boltzmann framework, 
the complete thermodynamics of counter-ions inside 
a charged sphere (salt-free system). 
\end{abstract}


\date{\today} 

\maketitle

\section{Introduction}
In the 1920's, Debye and H\"uckel (DH) \cite{Debye23} proposed a linearized 
mean-field description of the bulk thermodynamics of Coulomb
fluids, which works in the high-temperature region.
A few years earlier, Gouy \cite{Gouy10} and Chapman \cite{Chapman13} had
introduced the nonlinear Poisson-Boltzmann (PB) mean-field treatment
of the electric double layer. It paved the way toward the celebrated DVLO
theory of colloidal interactions \cite{Verwey48}.
When it comes to studying colloidal suspensions at finite density,
an efficient tool is furthermore provided by the cell model 
\cite{Belloni98,Hansen00,Lukatsky01,Levin02} in which the space around 
a large charged colloid is modelled by a spherical domain confining 
the mobile counter-ions of opposite charge.   
In the context of the PB cell model, the concept of (effective) charge
renormalization was introduced by Alexander et al \cite{Alexander84}.  

In the high-temperature region, the PB theory describes adequately many 
equilibrium and electrokinetic phenomena in Coulomb theory of neutral 
systems with repulsive and attractive forces among the charged objects.
Rigorous results on the existence and uniqueness of the solutions
of the PB equation were derived by mathematicians \cite{Hess73,Li09}. 
Two basic kinds of Coulomb systems are studied: the two-component electrolyte
of $\pm$ charges and the one-component systems of identical charges
with a neutralizing uniform background, distributed either in the domain's 
volume (jellium models) or on the domain's boundary 
(models with ``counter-ions only'').  
In the case of one-component systems, the PB equation belongs to
Liouville's type of non-linear differential equations.
Exact solutions are available for the effectively one-dimensional (1D) geometry
of parallel plates \cite{Andelman} and for the two-dimensional (2D) cylinder 
geometry \cite{Fuoss51}.
The latter solution is important in the Manning theory of counter-ions
condensation \cite{Manning69,TrTe06} which assumes that counter-ions can 
condense onto the polyion (a chain of monomer charges, often represented as 
an idealized line charge) up to a certain critical value.
The number density of counter-ions at the charged planar surface fulfills 
the so-called contact theorem
\cite{Henderson78,Henderson79,Choquard80,Carnie81,Totsuji81,Wennerstrom82}. 
An attempt to generalize the contact theorem to curved boundaries
was made recently in Ref. \cite{Mallarino14}.

In the case of purely attractive forces, the second-order Liouville equation 
plays a fundamental role in the gravitational theory of stellar structure 
\cite{Chandrasekhar67}, in diffusion in chemical kinetics \cite{Frank55}  
and in the theory of combustion and thermal explosion \cite{Zeldovich85}.
In contrast to Coulomb fluids, the Liouville equation of such systems 
exhibits minus sign ahead of the exponential (see below). 
For both Dirichlet and Neumann boundary conditions in the spherical geometry,
it exhibits multiplicity of solutions \cite{Steggerda65}.
The Liouville equation can be reduced to the first-order Abel's differential 
equation of the second kind, the so-called Enig's equation 
\cite{Enig67,Adler11}.

Developments of the Liouville equation in the theory of gravitational matter 
and related combustion systems were generally ignored by the Coulomb community 
because of its different layout and fundamental properties.
Only in a very recent study of the relaxation and the steady state 
with an initial injection of ions into a ball described by the 
Poisson-Nernst-Planck equations \cite{Schuss15}, was the PB equation with 
a specific initial value problem studied, predominantly numerically, 
by using equations of Enig's type.

In this work, we study a system of identical mobile point-like charges 
(counter-ions) confined to a domain with curved hard walls carrying 
a uniform fixed surface charge density, with the condition of
overall electroneutrality.
Three domain geometries are considered: a pair of parallel plates, 
the cylinder and the sphere.
The particles are in thermal equilibrium, and the nonlinear Poisson-Boltzmann 
theory rules the mean potential, with appropriate boundary conditions. 
While the effectively 1D parallel plates and the 2D cylinder have already 
been solved, the three-dimensional (3D) sphere problem has not.
The contact density of particles at the charged surface is shown to be
determined by a first-order Abel differential equation of the second kind, 
which is a counterpart of Enig's equation. 
This equation enables us to construct the exact series expansions of 
the contact density in the regions of small and large surface charge densities.
The formalism provides the complete thermodynamics of counter-ions inside 
the sphere with charged boundaries, within the PB framework.

The article is organized as follows.
In Sec. \ref{section:PB}, we introduce the  models studied and 
the PB formalism.
Sec. \ref{section:exact} brings an analysis of the solvable 1D parallel 
plates and the 2D cylindrical geometry.
Sec. \ref{section:Abel} is devoted to the derivation of the Abel differential
equation for a function which determines the contact density of counter-ions 
at the charged wall.
Sec. \ref{section:thermodynamics} deals with the application of the formalism
to the spherical geometry, which is not integrable.
The Abel equation enables us to construct the exact series expansions of 
the contact density in the regions of small and large surface charge densities.
Specific algebraic operations with the PB equation imply the corresponding 
series expansions for the internal energy and the free energy.
A brief recapitulation and concluding remarks are given in
Sec. \ref{section:conclusion}.

We emphasize that the geometries considered here for the cylindrical 
and spherical cases are somewhat simplified as compared to the 
widely used cell model for colloids. In the latter case indeed,
a rod-like or spherical charged macromolecule is placed at the
center of a concentric Wigner-Seitz cell \cite{Levin02,Fuoss51},
assumed to bear a vanishing charge. The PB problem should thus be
solved between two concentric bodies. Here, the inner body is
absent. The present case thus applies to confined situations,
such as ions in a charged nanotube or pore \cite{Akoum,Rotenberg} or in a charged spherical capsule \cite{Tsao}.

\section{Poisson-Boltzmann formalism} \label{section:PB}

\subsection{Studied models}
We consider a system of $N$ identical point-like particles of charge $-e$
(say $e$ is the elementary charge) immersed in a solvent which is considered 
as a medium of uniform dielectric permittivity $\varepsilon$ (in Gauss units).
The particles are confined to a 3D domain by hard walls.
For simplicity, the dielectric permittivity of the walls $\varepsilon'$ 
is the same as that of the solvent, $\varepsilon'=\varepsilon$, 
so the electrostatic image charges are absent.
The 2D domain surface carries a uniform surface charge 
density $\sigma e$, $\sigma>0$ having dimension $[{\rm length}]^{-2}$.
The system as a whole is electroneutral.
Since the charge of particles is opposite to the surface charge density,
they are coined as ``counter-ions''.
We consider three types of confining domain: 
two parallel plates forming a slab, the cylinder and the sphere.

The particles interact pair-wisely via the 3D Coulomb potential 
$v({\bf r}) = 1/(\varepsilon r)$ $(r\equiv\vert {\bf r}\vert)$.
It is the solution of the 3D Poisson equation
\begin{equation} \label{pointCoulomb}
\Delta v({\bf r}) = - \frac{4\pi}{\varepsilon} \delta({\bf r}) , 
\end{equation}
where $\delta$ is the Dirac delta function.
The particle system is in thermal equilibrium at the inverse temperature
$\beta=1/(k_{\rm B}T)$, where $k_{\rm B}$ is the Boltzmann constant.
It is useful to introduce the so-called Bjerrum length
\begin{equation}
\ell_{\rm B} \equiv \frac{\beta e^2}{\varepsilon} ,
\end{equation}
i.e. the distance at which two elementary charges in a solvent of
dielectric permittivity $\varepsilon$ interact with thermal energy $kT$.
Denoting by $\langle \cdots \rangle$ the statistical average over the
canonical ensemble, the particle number density $n({\bf r})$ and the charge 
density $\rho({\bf r})$ at point ${\bf r}$ are defined by
\begin{equation}
n({\bf r}) = \left\langle \sum_{i=1}^N \delta({\bf r}-{\bf r}_i) 
\right\rangle , \qquad \rho({\bf r}) = - e n({\bf r}) .
\end{equation}

For a given profile of the charge density $\rho({\bf r})$, the averaged
electric potential $\psi({\bf r})$ is given by the Poisson equation
\begin{equation} \label{Poisson}
\Delta \psi({\bf r}) = - \frac{4\pi}{\varepsilon} \rho({\bf r}) .
\end{equation}
For every geometry, the condition of overall electroneutrality is 
equivalent to specific Neumann boundary conditions (BCs) for 
the derivatives of $\psi({\bf r})$.
\begin{itemize}
\item
{\bf Parallel plates:} We consider that particles are confined to the
space between two infinite parallel plates at distance $R$.
The 2D surface at $r=0$ is not charged, the other one at $r=R$
is charged by the uniform surface charge density $\sigma e$.
For this effectively 1D geometry, the Laplacian
$\Delta\to {\rm d}^2/{\rm d}r^2$ and the Poisson equation takes the form
\begin{equation}
\frac{{\rm d}^2\psi(r)}{d r^2} = - \frac{4\pi}{\varepsilon} \rho(r) .
\end{equation}
Integrating this equation over $r$ from 0 to $R$, we obtain
\begin{equation}
\psi'(R) - \psi'(0) = - \frac{4\pi}{\varepsilon} \int_0^R {\rm d}r\, \rho(r) .
\end{equation}
The surface at $r=0$ is not charged which implies $\psi'(0)=0$.
The derivative of the electric potential at the charged surface is
determined by the electroneutrality condition 
$\int_0^R {\rm d}r\, \rho(r) + \sigma e = 0$.
We conclude that
\begin{equation} \label{BC0}
\psi'(0) = 0 , \qquad \psi'(R) = \frac{4\pi\sigma e}{\varepsilon} .
\end{equation}
\item
{\bf Cylindrical geometry:}
The next geometry corresponds to an infinitely long cylinder with radius $R$, 
whose surface carries the linear charge density $\lambda e$ (dimension of 
$\lambda$ is $[{\rm length}]^{-1}$) along the cylinder axis; 
$\lambda e$ is expressible in terms of the surface charge density $\sigma e$ 
as follows 
\begin{equation} \label{lambda}
\lambda e = 2\pi R \, \sigma e .
\end{equation}
Let us consider a plane perpendicular to the cylinder axis and denote
by $r$ $(0\le r\le R)$ the radial distance from this axis.
Because of the circular symmetry of the problem, the Laplacian takes the form
\begin{equation}
\Delta \to \frac{1}{r} \frac{{\rm d}}{{\rm d}r} \left( r 
\frac{{\rm d}}{{\rm d}r} \right) = \frac{{\rm d}^2}{{\rm d}r^2}
+ \frac{1}{r} \frac{{\rm d}}{{\rm d}r} .
\end{equation}
The electroneutrality condition 
$\int_0^R {\rm d}r\, 2\pi r \rho(r) + \lambda e = 0$, 
when combined with the Poisson equation 
\begin{equation}
\frac{1}{r} \frac{{\rm d}}{{\rm d}r} \left( r \frac{{\rm d}\psi(r)}{{\rm d}r} 
\right) = - \frac{4\pi}{\varepsilon} \rho(r) ,
\end{equation}
is equivalent to the BCs
\begin{equation}
\lim_{r\to 0} r \psi'(r) = 0 , \qquad 
R \psi'(R) = \frac{2\lambda e}{\varepsilon} .
\end{equation}
With regard to the relation (\ref{lambda}) between $\lambda$ and $\sigma$, 
the previous planar BC (\ref{BC0}) for $\psi'(R)$ is recovered.
\item
{\bf Spherical geometry:}
We consider a sphere of radius $R$ whose center is localized at 
the origin ${\bf 0}$.
The sphere surface carries the charge density $Z e$ (the valence $Z$ 
has dimension $[{\rm length}]^0$) where
\begin{equation} \label{Z}
Z = 4\pi \sigma R^2 .
\end{equation}
Because of the radial symmetry of the problem, the Laplacian reads as
\begin{equation}
\Delta \to \frac{1}{r^2} \frac{{\rm d}}{{\rm d}r} \left( r^2 
\frac{{\rm d}}{{\rm d}r} \right) = \frac{{\rm d}^2}{{\rm d}r^2}
+ \frac{2}{r} \frac{{\rm d}}{{\rm d}r} .
\end{equation}
The electroneutrality condition 
$\int_0^R {\rm d}r\, 4\pi r^2 \rho(r) + Z e = 0$, 
when combined with the Poisson equation 
\begin{equation}
\frac{1}{r^2} \frac{{\rm d}}{{\rm d}r} \left( r^2 
\frac{{\rm d}\psi(r)}{{\rm d}r} \right) = - \frac{4\pi}{\varepsilon} \rho(r) ,
\end{equation}
is equivalent to the BCs
\begin{equation}
\lim_{r\to 0} r^2 \psi'(r) = 0 , \qquad R^2 \psi'(R) = \frac{Z e}{\varepsilon} .
\end{equation}
The relation between $Z$ and $\sigma$ (\ref{Z}) implies again
the previous planar BC (\ref{BC0}) for $\psi'(R)$.
\end{itemize}

\subsection{The PB equation}
In the high-temperature limit and/or in the regions where the electrostatic 
potential is small, the statistical mechanics of our model is reasonably 
described by the PB approximation based on the mean-field assumption 
that the density of particles at point ${\bf r}$ is proportional to 
the Boltzmann factor taken with the energy $-e \psi({\bf r})$ of
the charge $-e$ in the mean electric potential $\psi({\bf r})$, i.e.
\begin{equation}
n({\bf r}) = n_0 \exp\left[ \beta e \psi({\bf r}) \right] ,
\end{equation}
where $n_0$ is a normalization factor.
Inserting this assumption into the Poisson equation (\ref{Poisson}) 
with $\rho({\bf r}) = -e n({\bf r})$, we get in terms of the reduced 
(dimensionless) potential $\phi({\bf r}) \equiv \beta e \psi({\bf r})$ 
\begin{equation} \label{Poisson2}
\Delta \phi({\bf r}) = 4\pi \ell_{\rm B} n_0 \, {\rm e}^{\phi({\bf r})} .
\end{equation}
Instead of the normalization factor $n_0$, we make use of the dimensionless 
quantity $z$ via $4\pi \ell_{\rm B} n_0 = z/R^2$.
We also introduce the geometry parameter $\alpha$, which is equal to: 
$0$ for the planar case, $1$ for the cylindrical geometry and $2$ for 
the spherical geometry.
Then Eq. (\ref{Poisson2}) can be written as
\begin{equation} \label{Poisson3}
\frac{{\rm d}^2 \phi(r)}{{\rm d}r^2} + 
\frac{\alpha}{r} \frac{{\rm d}\phi(r)}{{\rm d}r} 
\, = \, \frac{z}{R^2} \,{\rm e}^{\phi(r)} .
\end{equation} 
The BCs for the reduced electric potential read
\begin{equation} \label{BC1}
\lim_{r\to 0} r^{\alpha} \phi'(r) = 0 , \qquad R \phi'(R) = \eta ,
\end{equation}
where we introduced the dimensionless charge
\begin{equation}
\eta = 4\pi \ell_{\rm B} \sigma R .
\end{equation}
The profile of the particle number density is expressible as
\begin{equation}
n(r) = \frac{z}{4 \pi \ell_{\rm B} R^2}\, {\rm e}^{\phi(r)} .
\end{equation}

There is a gauge freedom in the PB equation (\ref{Poisson3}): 
the potential $\phi(r)$ is determined up to a constant,
which trivially renormalizes
$z$. 
We make the choice
\begin{equation}
\phi(R) = 0 ,
\end{equation}
which is merely dictated by convenience. All choices lead to 
equivalent descriptions (i.e. identical oservables such as densities).
Once a gauge has been chosen, for all three $\alpha=0,1,2$ geometries, 
the relation $z(\eta)$ between the dimensionless parameters $z$ and $\eta$ 
is fixed uniquely.
Note that in order to simplify the notation, we omit the dependence of 
$z(\eta)$ on the geometry parameter $\alpha$.
Having this relation available explicitly, the contact number density at 
the charged wall is given by
\begin{equation} \label{contactdensity}
n(R) = \frac{z(\eta)}{4 \pi \ell_{\rm B} R^2} .
\end{equation}
It will be shown later that for the spherical $\alpha=2$ geometry, 
the crucial function $z(\eta)$ determines not only the contact particle number
density, but also the complete thermodynamics of the counter-ion system. 
We finally emphasize that upon replacing $z$ by its expression (\ref{contactdensity}),
then the relations $z(\eta)$ to be reported are fully gauge-invariant.

The special case $\eta=0$ corresponds to $\sigma=0$, i.e. there are
no particles in the domain.
Consequently, we must have
\begin{equation} \label{BCz}
z(\eta=0) = 0 .
\end{equation}

\section{Exact solutions} \label{section:exact}
We now present the exact solutions of the PB equation for counter-ions 
between parallel plates (see also \cite{Andelman}) and in the cylinder
(see also \cite{Fuoss51}).
The way in which these exact solutions are formulated and derived will
impinge on the treatment of the spherical geometry.

\subsection{Parallel plates}
For two parallel plates at $r=0$ and $r=R$, the $\alpha=0$ PB equation
\begin{equation} \label{BP0}
\frac{{\rm d}^2 \phi(r)}{{\rm d}r^2} = \frac{z}{R^2} {\rm e}^{\phi(r)} ,
\end{equation} 
multiplied by $\phi'(r)$, can be integrated:
\begin{equation} \label{int}
\frac{1}{2} \left[ \phi'(r) \right]^2 = \frac{z}{R^2} 
\left[ {\rm e}^{\phi(r)} - c \right] .
\end{equation}
The integration constant $c$ is determined by the BC (\ref{BC1}) at $r=0$: 
$c=\exp[\phi(0)]$.
Taking into account that $\phi'(r)\ge 0$, Eq. (\ref{int}) is solved by
the method of the separation of variables, with the result
\begin{equation} \label{result}
\phi(r) = \ln c - 2 \ln \cos\left( \sqrt{\frac{cz}{2}}\frac{r}{R} \right) .
\end{equation}  
The BC (\ref{BC1}) at $r=R$ implies that
\begin{equation}
\eta = \sqrt{2cz} \tan\left( \sqrt{\frac{cz}{2}}\right) . 
\end{equation}
Fixing the gauge $\phi(R)=0$ leads to the condition
\begin{equation}
c = \cos^2\left( \sqrt{\frac{cz}{2}} \right) 
= 1 - \frac{\eta^2}{2z} .
\end{equation} 
Finally, we arrive at the implicit relation between $z$ and $\eta$:
\begin{equation} \label{zeta0}
\eta = \sqrt{2z} \sin\left( \sqrt{\frac{z}{2}-\frac{\eta^2}{4}} \right) .
\end{equation}

Eq. (\ref{zeta0}) provides the explicit forms of the series expansions of
$z(\eta)$ for small values of $\eta$ (small surface charge density and/or
distance between the plates) and large $\eta$ (large surface charge density 
and/or distance between the plates).
\begin{itemize}
\item
The small-$\eta$ series expansion turns out to be analytic and reads
\begin{eqnarray} 
z & = & \eta + \frac{\eta^2}{3} + \frac{\eta^3}{45} 
- \frac{2\eta^4}{945} + \frac{\eta^5}{14175} \nonumber \\ & & 
+ \frac{2\eta^6}{93555} - \frac{1082\eta^7}{212837625} + \cdots .
\label{smalleta0}
\end{eqnarray}
This expansion holds in the case of weak potentials for which the exponential
$\exp[\phi(r)]$ in the PB equation (\ref{BP0}) can be systematically expanded 
in powers of $\phi(r)$.
The leading term in $\eta$ corresponds to the Debye-H\"uckel (DH) approximation
which arises from the lowest-order expansion $\exp[\phi(r)]\sim 1$. 
The next term $\eta^2/3$ has its origin in the series expansion of 
the exponential up to the term $\phi(r)/1!$, and so on.
In the Appendix, we derive the first two terms of the small-$\eta$ 
expansion of $z$ for all three $\alpha=0,1,2$ geometries.
For $\alpha=0$, the obtained result (\ref{fexp}) agrees with the expansion 
(\ref{smalleta0}). 
\item
The large-$\eta$ series expansion is more delicate.
In the limit $\eta\to\infty$, we find that $cz=\pi^2/2$.
Since $cz=z-\eta^2/2$, the first two expansion terms read as
$z=\eta^2/2+\pi^2/2$.
We can proceed further to obtain the non-analytic expansion
\begin{eqnarray}
z & = & \frac{\eta^2}{2} + \frac{\pi^2}{2} - \frac{2\pi^2}{\eta}
+ \frac{6\pi^2}{\eta^2} - \frac{2\pi^2 (24-\pi^2)}{3\eta^3}  
\nonumber \\ & & - \frac{20\pi^2 (\pi^2-6)}{3\eta^4} + \cdots .
\label{largeeta0}
\end{eqnarray}
This expansion was derived using the fact that $cz$ goes to its 
asymptotic value $\pi^2/2$ from below.
The leading term $\eta^2/2$ is expected, since for asymptotically large 
distances between the plates, the density of particles at the single charged 
planar wall is fixed by the contact theorem 
\cite{Henderson78,Henderson79,Choquard80,Carnie81,Totsuji81,Wennerstrom82} 
to the value
\begin{equation} 
\lim_{R\to\infty} n(R) = 2\pi \ell_{\rm B} \sigma^2 =
\frac{\eta^2/2}{4\pi \ell_{\rm B} R^2} .
\end{equation}
In view of the formula for the contact density (\ref{contactdensity}),
this means that $z\sim \eta^2/2$.  
As we shall see, the same leading term $z\sim\eta^2/2$ is present for all 
three $\alpha=0,1,2$ geometries. Indeed, $\eta\to\infty$ can be envisioned as
the planar limit of zero surface curvature, where furthermore the inter-plate
distance is divergent.
\end{itemize}

Differentiating both sides of (\ref{zeta0}) with respect to $\eta$,
it is straightforward to derive the first-order nonlinear differential 
equation for the function $z(\eta)$:
\begin{equation} \label{dzeta0}
\frac{{\rm d}z}{{\rm d}\eta} = \frac{(2+\eta) z}{z+\eta} , \qquad
z(\eta=0) = 0 .
\end{equation}
This equation is of Abel's type and belongs to integrable differential
equations (see e.g. \cite{Kamke}).
The small-$\eta$ series expansion of $z$ (\ref{smalleta0}) can be derived
trivially from (\ref{dzeta0}).
On the other hand, the large-$\eta$ expansion of $z$ (\ref{largeeta0})
is determined by (\ref{dzeta0}), up to the constant term $\pi^2/2$.
The point is that the BC $z(\eta=0)=0$ complements the differential 
equation (\ref{dzeta0}) in the region of small $\eta$'s, while for 
large $\eta$'s, an integration constant is missing.
Would the analytical solution not be available, 
the missing constant term $\pi^2/2$ could be deduced with a high precision 
e.g. by solving the differential equation (\ref{dzeta0}) numerically, 
going from small to large $\eta$'s, and then subtracting 
the known leading large-$\eta$ term $\eta^2/2$.

The contact theorem for the particle densities at planar plates, 
when applied to our model, states that
\begin{equation}
n(0) = \beta P , \qquad n(R) - n(0) = 2\pi \ell_{\rm B} \sigma^2 ,
\end{equation}
where $P$ is the pressure.
The explicit results
\begin{equation}
n(0) = \frac{1}{4\pi \ell_{\rm B}R^2} \left( z - \frac{\eta^2}{2} \right) ,
\qquad n(R) = \frac{z}{4\pi \ell_{\rm B}R^2} 
\end{equation}
agree with the value of the density difference and set
\begin{equation}
\beta P = \frac{1}{4\pi \ell_{\rm B}R^2} \left( z - \frac{\eta^2}{2} \right) .
\end{equation}
For small $\eta$, $\beta P$ admits the series expansion
\begin{equation}
\beta P = \frac{1}{4\pi \ell_{\rm B}R^2} \left( \eta - \frac{\eta^2}{6} 
+ \frac{\eta^3}{45} - \frac{2 \eta^4}{945} + \cdots \right) .
\end{equation}
For $R\to 0$, $\beta P$ diverges as $\sigma/R$.
In addition, the large-$\eta$ expansion of the pressure reads as
\begin{equation}
\beta P = \frac{1}{4\pi \ell_{\rm B}R^2} \left( \frac{\pi^2}{2} 
- \frac{2\pi^2}{\eta} + \frac{6\pi^2}{\eta^2} + \cdots \right) .
\end{equation}
In the limit $R\to\infty$, $\beta P$ goes to 0 like $\pi/(8\ell_{\rm B}R^2)$,
a universal expression independent of the bare surface charge density. 
This is an illustration of the saturation phenomenon \cite{TeTr03},
central to the physics of effective charges in colloidal suspensions 
\cite{BoTA02}.

The internal energy $U$ of the interacting charges is contained in 
the electric field as follows
\begin{equation}
\frac{\beta U}{\Sigma} = \frac{\beta \varepsilon}{8\pi} \int_0^R {\rm d}r\, 
[\psi'(r)]^2 = \frac{1}{8\pi \ell_{\rm B}} \int_0^R {\rm d}r\, [\phi'(r)]^2 ,
\end{equation}
where $\Sigma$ is the (infinite) surface of either of the walls.
Analogously, we have \cite{Trizac96}
\begin{eqnarray}
\frac{\beta U}{\Sigma} & = & \frac{1}{2} \int_0^R {\rm d}r\, 
\left[ \sigma e \delta(r-R) - e n(r) \right] \beta\psi(r) \nonumber \\
& = & - \frac{1}{2} \int_0^R {\rm d}r\, n(r) \phi(r) . 
\end{eqnarray}
Using the result (\ref{result}) and the fact that the particle
number $N=\sigma\Sigma$, we obtain
\begin{equation}
\frac{\beta U}{N} = 1 - \frac{1}{\eta} \left( z - \frac{\eta^2}{2} \right) .
\end{equation} 

To obtain the free energy $F = U-T S$ with $S$ being the entropy, 
we use the PB expression \cite{Trizac96}
\begin{equation}
\frac{S}{\Sigma} = - k_{\rm B} \int_0^R {\rm d}r\,
n(r) \left\{ \ln \left[ \Lambda^3 n(r) \right] - 1 \right\} ,
\end{equation}
where $\Lambda$ is the thermal de Broglie wavelength.
Simple algebra leads to
\begin{equation}
- \frac{\beta (TS)}{N} = - 2 \frac{\beta U}{N} + \ln\Lambda^3 - 1
+ \ln\left( \frac{z}{4\pi\ell_{\rm B}R^2} \right) .
\end{equation}
This formula implies that within the PB approximation
\begin{equation}
\frac{\beta F}{N} = - \frac{\beta U}{N} + \ln\Lambda^3 - 1
+ \ln\left( \frac{z}{4\pi\ell_{\rm B}R^2} \right) .
\end{equation}
For the ideal non-interacting gas system of volume $\Sigma R$, 
the free energy $F^{\rm id}$ is given by
\begin{equation}
\beta F^{\rm id} = \ln\left( N! \Lambda^{3N} \right) - N \ln(\Sigma R) .
\end{equation}
The excess (i.e. over ideal) free energy is defined as 
$F^{\rm ex}\equiv F - F^{\rm id}$.
Using Stirling's formula for $\ln(N!)$ and taking the thermodynamic
limit $N\to\infty$, we arrive at
\begin{equation}
\frac{\beta F^{\rm ex}}{N} = \frac{1}{\eta} \left( z - \frac{\eta^2}{2} \right) 
- 1 + \ln\left( \frac{z}{\eta} \right) .
\end{equation}
It is easy to verify that the thermodynamic relation
\begin{equation}
\beta P = - \frac{\partial}{\partial R} \left( \frac{\beta F}{\Sigma} \right) 
\end{equation}
holds.
For small $\eta$, the excess free energy exhibits the analytic expansion
\begin{equation}
\frac{\beta F^{\rm ex}}{N} = \frac{\eta}{6} - \frac{\eta^2}{90} +
\frac{2\eta^3}{2835} - \frac{\eta^4}{56700} - \frac{2\eta^5}{467775} +
\cdots .
\end{equation}
Note that $F^{\rm ex}$ vanishes in the non-interacting limit $\eta\to 0$, 
as it should be.
In the large-$\eta$ region, the expansion of the excess free energy reads as
\begin{equation}
\frac{\beta F^{\rm ex}}{N} = \ln\eta - (1+\ln 2) + \frac{\pi^2}{2\eta}
- \frac{\pi^2}{\eta^2} + \frac{2\pi^2}{\eta^3} + \cdots .
\end{equation}

\subsection{Cylindrical geometry}
The PB equation for the $\alpha=1$ cylindrical geometry 
\begin{equation} 
\frac{{\rm d}^2 \phi(r)}{{\rm d}r^2} + 
\frac{1}{r} \frac{{\rm d}\phi(r)}{{\rm d}r} 
= \frac{z}{R^2} {\rm e}^{\phi(r)} ,
\end{equation} 
complemented by the BC $\lim_{r\to 0} r\phi'(r) = 0$, provides the solution
\begin{equation}
\phi(r) = - 2 \ln (b^2-r^2) + \ln (8 b^2) - \ln\left( \frac{z}{R^2} \right) .
\end{equation}
The parameter $b$ is determined by the BC $R\phi'(R)=\eta$ as follows
\begin{equation}
b^2 = R^2 \left( 1 + \frac{4}{\eta} \right) .
\end{equation}
The gauge $\phi(R)=0$ implies the relation between $z$ and $\eta$ of
the simple form
\begin{equation} \label{zeta1}
z = \frac{\eta^2}{2} + 2 \eta .
\end{equation}
Both terms are reproduced in the expansion around DH limit, see 
Eq. (\ref{fexp}) in the Appendix.
The particle number density is given by
\begin{equation}
n(r) = \frac{2 b^2}{\pi \ell_{\rm B}} \frac{1}{(b^2-r^2)^2} .
\end{equation}

It is interesting that the function $z(\eta)$ can be deduced directly from
the PB equation, written in the form
\begin{equation} \label{PBaux}
\frac{{\rm d}}{{\rm d}r} \left[ r\phi'(r) \right] = \frac{z r}{R^2}
{\rm e}^{\phi(r)} .
\end{equation}
Multiplying this equation by $r\phi'(r)$ and integrating over $r$
from $0$ to $R$, we get
\begin{eqnarray}
\frac{1}{2} \left[ R \phi'(R) \right]^2 & = & \frac{z}{R^2}
\int_0^R {\rm d}r\, r^2 \frac{{\rm d}}{{\rm d}r} {\rm e}^{\phi(r)} \nonumber \\
& = & \frac{z}{R^2} \left[ R^2 - 2 \int_0^{R} {\rm d}r\, r {\rm e}^{\phi(r)} 
\right] .
\end{eqnarray}
Using once more the PB equation (\ref{PBaux}), we arrive at (\ref{zeta1}).

The first-order differential equation following from (\ref{zeta1}) reads
\begin{equation} \label{dzeta1}
\frac{{\rm d}z}{{\rm d}\eta} = 2+\eta , \qquad
z(\eta=0) = 0 .
\end{equation}
It is trivially integrated, and there is thus a single expression $z(\eta)$
to cover both the regimes of small and large $\eta$.

The present geometry is convenient for deriving thermodynamic relations.
In order to obtain the internal energy, we invoke a detour,
which relies on the fact that the PB mean-field treatment is actually 
space-dimension independent.
This means that the PB formulation for two-dimensional charges or for 
three-dimensional charges in cylindrical confinement coincide. 
We thus address momentarily the 2D case, which can be envisioned
as dealing with a collection of parallel lines in 3D. 
The pairwise Coulomb interactions among particles are no longer 
$1/(\varepsilon r)$, but the effective ones given by the 2D version
of the generic Poisson Eq. (\ref{pointCoulomb}), 
$v(r)=-(2/\varepsilon) \ln r$.
The number of particles per unit length along the cylinder axis
\begin{equation}
N = \int_0^R {\rm d}r\, 2\pi r n(r) = 2\pi \sigma R
\end{equation} 
fulfills the 2D electroneutrality condition, $\sigma e$ being the
line charge density on the disk boundary.
The potential induced by the line charge density $\sigma e$ inside the
disc is constant, equal to $-(2/\varepsilon) N e \ln R$.
To obtain the internal energy of the Coulomb system, we have to sum 
the interaction energy of the fixed line charge density with itself,
\begin{equation}
E_{ss} = \frac{1}{2} N e \left( -\frac{2}{\varepsilon} N e \ln R \right) 
= - \frac{(N e)^2 \ln R}{\varepsilon} ,
\end{equation}
the interaction energy of the fixed line charge density with particles,
\begin{equation}
E_{sp} = (- N e) \left( -\frac{2}{\varepsilon} N e \ln R \right)  
= \frac{2 (N e)^2 \ln R}{\varepsilon} ,
\end{equation}
and finally the mean particle-particle interaction energy
\begin{equation} \label{energypp}
E_{pp} = \frac{1}{2} \int_0^R {\rm d}^2 r \int_0^R {\rm d}^2 r'\, n(r) 
\left( - \frac{2 e^2}{\varepsilon} \ln \vert {\bf r}-{\bf r}' \vert \right) 
n(r') .  
\end{equation}
The integral can be simplified by using the equality for two points
with polar coordinates ${\bf r}=(r,\varphi)$ and ${\bf r}'=(r',\varphi')$:
\begin{equation}
-\ln \vert {\bf r}-{\bf r}' \vert = - \ln r_{>} + \sum_{j=1}^{\infty}
\frac{1}{j} \left( \frac{r_{<}}{r_{>}} \right)^j
\cos j (\varphi'-\varphi) ,
\end{equation}
where $r_>=\max(r,r')$ and $r_<=\min(r,r')$.
The terms with positive $j$ do not contribute to the integral
in (\ref{energypp}) after integration over either angle $\varphi$
or angle $\varphi'$.
Taking advantage of the $(r,r')$ symmetry, we can write
\begin{equation} 
E_{pp} = - \frac{2 e^2}{\varepsilon} \int_0^R {\rm d}^2 r\, n(r) \ln r 
\int_0^r {\rm d}^2 r'\, n(r') .  
\end{equation}
Evaluating the integral, the total internal energy per unit length along
the cylinder axis $U = E_{ss} + E_{sp} + E_{pp}$ is given by
\begin{equation} \label{U}
\frac{\beta U}{N} = 1 - \frac{4}{\eta} \ln \left( 1 + \frac{\eta}{4} \right) .
\end{equation}
An alternative way to calculate $U$ is to use the formula 
\begin{equation}
\beta U = \frac{1}{8\pi \ell_{\rm B}} \int_0^R {\rm d}^2r\, [\phi'(r)]^2 .
\end{equation}
Computing the integral
\begin{eqnarray}
\int_0^R {\rm d}r\, 2\pi r [\phi'(r)]^2 & = & 32\pi \int_0^R {\rm d}r\,
\frac{r^3}{(b^2-r^2)^2} \nonumber \\ & = & 8\pi \left[ \frac{\eta}{2} - 
2 \ln\left( 1 + \frac{\eta}{4} \right) \right] ,
\end{eqnarray}
we recover the laboriously derived result (\ref{U}).

With the aid of the same procedure as in the plane geometry, 
the excess free energy per particle is found to be
\begin{equation} \label{expr}
\frac{\beta F^{\rm ex}}{N} = -1 + \left( 1 + \frac{4}{\eta} \right) 
\ln\left( 1 + \frac{\eta}{4} \right)  .
\end{equation}
The small-$\eta$ expansion of this expression reads as
\begin{equation}
\frac{\beta F^{\rm ex}}{N} = \frac{\eta}{8} - \frac{\eta^2}{96}
+ \frac{\eta^3}{768} - \frac{\eta^4}{5120} + \frac{\eta^5}{30720} 
+ \cdots .
\end{equation}
The large-$\eta$ expansion of (\ref{expr}) takes the form
\begin{eqnarray}
\frac{\beta F^{\rm ex}}{N} & = & \ln\eta - (1+\ln 4) + \frac{4\ln\eta}{\eta}
+ \frac{4(1-\ln4)}{\eta} \nonumber \\
& & + \frac{8}{\eta^2} - \frac{32}{3\eta^3} + \frac{64}{3\eta^4}
+ \cdots .
\end{eqnarray}

The pressure is not uniquely defined for our curved geometry.
In contrast to the standard jellium models in which the neutralizing
background charge fills uniformly the domain's volume, our system has 
the neutralizing surface charge density on the domain boundary.
It still suffers from the jellium-like problems when calculating the pressure.
According to the standard definition, the pressure is related to 
the derivative of the free energy with respect to the volume 
at the fixed number of particles.
Changing the volume/surface of the domain involves the change of 
the background charge which must be compensated by the change of
the particle number in order to maintain the overall electroneutrality.
We refer to the work of Choquard et al. \cite{Choquard80} for a discussion 
of the various definitions of the pressure in jellium-like systems.

\section{Derivation of Abel's equation for counter-ions} 
\label{section:Abel}
In this section, we aim at deriving a differential equation directly for 
the crucial function $z(\eta)$ for all three geometries.
We shall adapt the procedure for gravitational systems presented e.g. 
in Ref. \cite{Adler11}.

We first rewrite the PB equation (\ref{Poisson3}) into the form
\begin{equation} \label{PB}
\frac{1}{2} r \frac{{\rm d}}{{\rm d}r} 
\left( r \frac{{\rm d}\phi}{{\rm d}r} \right)
+ \frac{\alpha-1}{2} r \frac{{\rm d}\phi}{{\rm d}r}
= \frac{1}{2} \frac{z r^2}{R^2} {\rm e}^{\phi} .
\end{equation}
Let us define the new variable
\begin{equation}
\zeta \equiv \ln\left( \frac{2 R^2}{z r^2} \right) 
\end{equation}
and the functions
\begin{equation}
\theta(\zeta) \equiv \phi - \zeta , \qquad 
p(\zeta) \equiv \frac{{\rm d}\theta}{{\rm d}\zeta} , \qquad
q(\zeta) \equiv \frac{{\rm d}^2\theta}{{\rm d}\zeta^2} .
\end{equation}
Since $r {\rm d}_r = r ({\rm d}\zeta/{\rm d}r) d_{\zeta} = -2 {\rm d}_{\zeta}$,
the PB equation (\ref{PB}) is now equivalent to
\begin{equation}
2 q - (\alpha-1)(p+1) = {\rm e}^{\theta} .
\end{equation}
We introduce another function $Q$ via
\begin{equation} \label{Qq}
\theta = \ln(2 Q) , \qquad 
Q = q - \frac{(\alpha-1)(p+1)}{2} .
\end{equation}
Since it holds
\begin{equation}
p \equiv \frac{{\rm d}\theta}{{\rm d}\zeta} = 
\frac{1}{Q} \frac{{\rm d}Q}{{\rm d}\zeta} =
\frac{1}{Q} q \frac{{\rm d}Q}{{\rm d}p} ,
\end{equation}
expressing from (\ref{Qq}) $q$ in terms of $p$ and $Q$,
we obtain the linear differential equation for $Q$
as the function of $p$:
\begin{equation} \label{Abel1}
\left[ 2Q + (\alpha-1)(p+1) \right] \frac{{\rm d}Q}{{\rm d}p}
- 2 Q p = 0 .
\end{equation}

Now we return to the original $r$ variable and the electric potential
$\phi(r)$ and express in terms of them the new $p$ variable 
\begin{equation}
p(r) \equiv \frac{{\rm d}\theta}{{\rm d}\zeta} = 
\frac{{\rm d}\phi}{{\rm d}\zeta} - 1 = - \frac{1}{2} r\phi'(r) - 1 
\end{equation}
and the $Q$ function
\begin{equation}
Q(r) \equiv \frac{1}{2} {\rm e}^{\theta} = \frac{1}{2} {\rm e}^{\phi-\zeta}
= \frac{z r^2}{4 R^2} {\rm e}^{\phi(r)} . 
\end{equation}
In view of the last two relations, the BC at $r=0$ corresponds to 
$p=-1$ and $Q=0$ which is fully consistent with Eq. (\ref{Abel1}).
Under the gauge $\phi(R)=0$, the BC at $r=R$ corresponds to
\begin{equation}
p = -\frac{1}{2} \eta - 1 , \qquad Q = \frac{z}{4} .
\end{equation}
Inserting these relations into Eq. (\ref{Abel1}) we end up with
the first-order Abel differential equation of the second kind
\begin{equation} \label{finalde}
\frac{{\rm d}z}{{\rm d}\eta} = \frac{(2+\eta)z}{z-(\alpha-1)\eta} ,
\qquad z(\eta=0) = 0 .
\end{equation}
For $\alpha=0$ and 1, this equation coincides with the exact
ones (\ref{dzeta0}) and (\ref{dzeta1}), respectively.

Although our Coulomb Eq. (\ref{finalde}) was derived in analogy with 
gravitational Enig's equation \cite{Enig67,Adler11}, it differs 
fundamentally from the latter one.
Like for instance, the multiplicity of solutions of Enig's equation 
for the spherical geometry \cite{Steggerda65} is absent in its Coulomb
counterpart (\ref{finalde}).
Eq. (\ref{finalde}) provides the analytic series expansion of $z(\eta)$ 
for small values of $\eta$:
\begin{eqnarray}
z & = & (1+\alpha) \eta + \frac{1+\alpha}{3+\alpha} \eta^2 +
+ \frac{(1+\alpha)(1-\alpha)}{(3+\alpha)^2(5+\alpha)} \eta^3 \nonumber \\ 
& & - \frac{2(1+\alpha)(1+2\alpha)(1-\alpha)}{(3+\alpha)^3(5+\alpha)(7+\alpha)} 
\eta^4 + {\cal O}(\eta^5) .
\end{eqnarray}
The first two terms are in agreement with those obtained by the systematic 
expansion around the linear DH limit, see formula (\ref{fexp}) in the Appendix.
In the limit of large $\eta$, the differential Eq. (\ref{finalde}) implies
the non-analytic expansion 
\begin{eqnarray}
z & = & \frac{1}{2} \eta^2 + 2\alpha \eta + 4\alpha(1-\alpha) \ln\eta + c
- 16\alpha (1-\alpha)^2 \frac{\ln \eta}{\eta} \nonumber \\ & & 
+ 4 (1-\alpha)[2\alpha(3\alpha-1)-c] \frac{1}{\eta} 
+ {\cal O}\left( \frac{\ln\eta}{\eta^2}\right) , 
\end{eqnarray}
where the integration constant $c\equiv c(\alpha)$ depends on the geometry. 
The higher-order terms are of the form $(\ln\eta)^j/\eta^k$ 
with positive integers $j\le k$.
In the leading order, we recover the quasi-planar term $\eta^2/2$ 
independent of the geometry $\alpha$.
As was shown in the previous two sections, $c$ is equal to $\pi^2/2$ 
for $\alpha=0$ and to $0$ for $\alpha=1$.
This constant is unknown for the spherical $\alpha=2$ geometry and it
has to be determined, at least approximately, in an alternative way.

\section{Contact density and thermodynamics for 
spherical geometry} \label{section:thermodynamics}
\begin{figure}[htb]
\begin{center}
\includegraphics[width=0.45\textwidth,clip]{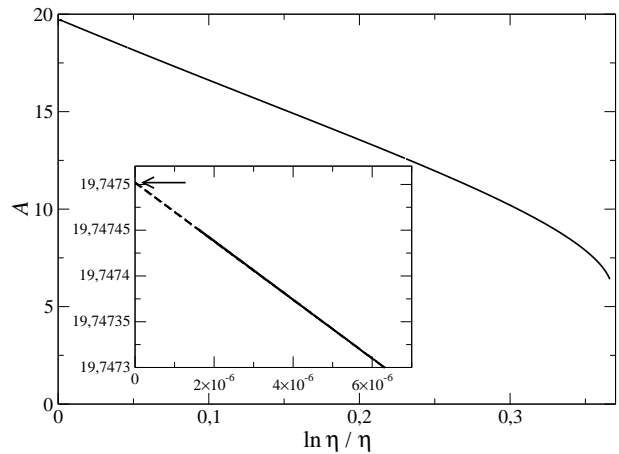}
\caption{Determination of the integration constant $c\equiv c(\alpha=2)$ for 
the spherical geometry. 
The differential equation (\ref{eq:diffsphere}) is solved numerically.
Use is then made of the large $\eta$ expansion (\ref{large2}). 
By plotting $A = z-\eta^2/2-4\eta +8 \ln\eta$ as a function of 
$(\ln\eta)/\eta$, we expect a linear behavior in the vicinity of the origin,
the extrapolation of which at $(\ln\eta)/\eta\to 0$ yields $c$. 
The main graph corresponds to the range $3<\eta<5\,10^6$ while the inset is 
a zoom in the upper-left corner, corresponding to the large $\eta$ regime, 
where the expected linear behavior is met. 
The dashed line shows the best fit to the numerical data,
with slope 32 as predicted by Eq. (\ref{large2}).
The extrapolation yields $c$, shown with an arrow,
with the value $c \simeq 19.747502$.}
\label{fig:c} 
\end{center}
\end{figure}

The PB equation for the spherical geometry ($\alpha=2$) reads 
\begin{equation} \label{PB2}
\frac{{\rm d}^2 \phi(r)}{{\rm d}r^2} + 
\frac{2}{r} \frac{{\rm d}\phi(r)}{{\rm d}r} 
= \frac{z}{R^2} {\rm e}^{\phi(r)} .
\end{equation} 
It is complemented by the BCs $\lim_{r\to 0} r^2\phi'(r) = 0$ 
and $R\phi'(R)=\eta$.
The corresponding differential equation for $z(\eta)$ is
\begin{equation} 
\frac{{\rm d}z}{{\rm d}\eta} = \frac{(2+\eta)z}{z-\eta} ,
\qquad z(\eta=0) = 0 .
\label{eq:diffsphere}
\end{equation}
As emphasized above, this Abel equation does not belong to 
the integrable ones \cite{Kamke}.
The corresponding small-$\eta$ series expansion is
\begin{eqnarray}
z & = & 3\eta + \frac{3}{5} \eta^2 - \frac{3}{175} \eta^3 
+ \frac{2}{525} \eta^4 - \frac{991}{1010625} \eta^5 \nonumber \\
& & + \frac{18166}{65690625} \eta^6 + {\cal O}(\eta^7) . \label{small2}
\end{eqnarray}
It is straightforward to generate also the higher-order terms 
of this series expansion, making use of a symbolic computation software
(to get the series
up to the term $\eta^{20}$, it requires on the order of a second of CPU on a standard modern
computer).
The series representation works well up to $\eta\simeq 2$.
On the other hand, the large-$\eta$ expansion is of the form
\begin{eqnarray}
z & = & \frac{1}{2} \, \eta^2 + 4 \,\eta - 8 \ln\eta + c
- 32 \,\frac{\ln \eta}{\eta} + 4 (c-20) \,\frac{1}{\eta} \nonumber \\ 
& & + 96 \frac{\ln\eta}{\eta^2} + 4(8-3c) \frac{1}{\eta^2} + \cdots .
\label{large2}
\end{eqnarray}
By comparison with the large-$\eta$ expansion for
parallel plates, Eq. (\ref{largeeta0}), we see that the curvature effect of
the sphere surface is embodied in the linear and logarithmic terms. 
Although there exists an implicit solution of Abel's equations including 
the non-integrable ones \cite{Panayotounakos05,Panayotounakos11},
we failed in deducing from it the integration constant $c$.  
Our numerical estimate is $c\simeq 19.747502$,
see Fig. \ref{fig:c}. In the same spirit, is possible to check the consistency 
of this estimate and of expansion (\ref{large2}) by plotting 
$\eta[z-\eta^2/2-4\eta +8 \ln\eta -c -32 (\ln\eta)/\eta ]$
as a function of $(\ln\eta)/\eta$. 
Extrapolating this quantity for $(\ln\eta)/\eta \to 0$ should give 
$4 (c-20)\simeq -1.01$. 
We have verified that this indeed is the case.

To obtain the thermodynamics of the spherical system, we return to 
the original PB equation (\ref{PB2}). 
When multiplied by $r$, it can be rewritten as
\begin{equation}
\frac{{\rm d}}{{\rm d}r} \left[ r \phi'(r) \right] + \phi'(r)
= \frac{z r}{R^2} {\rm e}^{\phi(r)} .
\end{equation}
We first multiply this equation by $r^2 \phi'(r)$ and then integrate
over $r$ from 0 to $R$, with the result
\begin{eqnarray}
\frac{1}{2} \int_0^R {\rm d}r\, r \frac{{\rm d}}{{\rm d}r}
\left[ r \phi'(r) \right]^2 + \int_0^R {\rm d}r\, \left[ r \phi'(r) \right]^2
\nonumber \\ = \frac{z}{R^2} \int_0^R {\rm d}r\, r^3 
\frac{{\rm d}}{{\rm d}r} {\rm e}^{\phi(r)} .
\end{eqnarray}
The integration by parts of the first integrals on the left and right sides 
leads to
\begin{equation} \label{prel}
\frac{1}{2} R \eta^2 + \frac{1}{2} \int_0^R {\rm d}r\, 
\left[ r \phi'(r) \right]^2 = \frac{z}{R^2} \left[ R^3
- 3 \int_0^R {\rm d}r\, r^2 {\rm e}^{\phi(r)} \right] .
\end{equation}
Rewriting equation (\ref{PB2}) as
\begin{equation}
\frac{{\rm d}}{{\rm d}r} \left[ r^2 \phi'(r) \right] =
\frac{z}{R^2} r^2 {\rm e}^{\phi(r)}
\end{equation} 
and integrating over $r$ from 0 to $R$, we find that
\begin{equation}
\frac{z}{R^2} \int_0^R {\rm d}r\, r^2 {\rm e}^{\phi(r)} = R\eta .
\end{equation}
Considering this equality in Eq. (\ref{prel}), we get
\begin{equation}
\frac{1}{2R} \int_0^R {\rm d}r\, r^2 \left[ \phi'(r) \right]^2
= z - \frac{1}{2} \eta^2 - 3 \eta . 
\end{equation}
Thus the (dimensionless) internal energy is given by
\begin{equation} \label{defU}
\beta U = \frac{1}{8\pi \ell_{\rm B}} \int_0^R {\rm d}r\, 4\pi r^2
\left[ \phi'(r) \right]^2 = \frac{R}{\ell_{\rm B}}
\left( z - \frac{1}{2} \eta^2 - 3 \eta \right) . 
\end{equation}
Since there are $N=4\pi R^2\sigma$ particles in the ball,
we have the relation for the internal energy per particle 
\begin{equation} \label{exactrel}
\frac{\beta U}{N} = \frac{1}{\eta}
\left( z - \frac{1}{2} \eta^2 - 3 \eta \right) . 
\end{equation}
It is interesting that this exact result follows from simple manipulations
with the PB equation, without solving explicitly the spherical system.
Since the series representations of $z$ in terms of $\eta$ are at our
disposal, this means the complete solution of thermodynamics for
counter-ions inside a sphere.

The excess free energy per particle is obtained in the form
\begin{equation}
\frac{\beta F^{\rm ex}}{N} = - \frac{1}{\eta} \left( 
z - \frac{\eta^2}{2} -3 \eta \right) + \ln\left( \frac{z}{3\eta} \right) .
\end{equation}  
In the limit of small $\eta$, this formula implies the expansion
\begin{equation}
\frac{\beta F^{\rm ex}}{N} = \frac{\eta}{10} - \frac{3\eta^2}{350} + 
\frac{2\eta^3}{1575} - \frac{991\eta^4}{4042500} 
+ \frac{18166\eta^5}{328453125} + \cdots .
\end{equation}
In the limit of large $\eta$, we find that
\begin{equation}
\frac{\beta F^{\rm ex}}{N} = \ln\eta - (1+\ln 6) + 8 \frac{\ln\eta}{\eta} 
+ \frac{8-c}{\eta} + \cdots .
\end{equation}

\section{Conclusion} \label{section:conclusion}
We have studied a system of identical counter-ions inside a homogeneously 
charged sphere surface, within the Poisson-Boltzmann mean-field theory.
For our salt-free system, there exist a clear-cut criterion
for the validity of Poisson-Boltzmann approach, as compared to 
an exact statistical mechanics solution, treating all charges as
interacting through the bare Coulomb potential, in a medium
of fixed dielectric permittivity (the so-called primitive model).
In the spherical geometry, the Coulombic coupling is measured 
by the coupling parameter $\Xi = 2 \pi \ell_B^2 \sigma$,
and the PB treatment is appropriate for $\Xi < 1$.
For larger coulings, non mean-field effects appear,
such as overcharging of like-charge attraction \cite{Levin02,Netz,Nagi,PRL11,Juan}.

Using techniques applied to Liouville equation,
we derived the PB-exact series expansions of the contact density and of 
the thermodynamics (the internal and free energies) in the regions of small 
and large surface charge densities/sphere radius. 
The derivation of the series expansions is
straightforward and very high orders can be readily obtained on the
order of a second of CPU.
As was indicated in the Appendix for the case of small surface charge 
densities, the systematic generation of the series by the standard expansion 
around the Debye-H\"uckel limit of weak charges is cumbersome, 
and one can reach with an increasing difficulty the first few terms only.
In the limit of large surface charge densities, one integration constant
is missing; it can be determined numerically with a high precision.

The cell model of colloidal suspensions requires to solve the PB equation for
counter-ions between two concentric spheres with charged surfaces.
This problem brings into the consideration two length scales, the radiuses 
of the inner and outer spheres, with the corresponding Neumann boundary 
conditions. 
It is not clear whether the  techniques presented here can be generalized 
to such a geometry of confinement.
If yes, previous simplified approximations for curved geometries, such as 
the application of the contact theorem valid for planar walls to the cell 
boundaries in \cite{Santos09}, might be replaced by rigorous approaches.     

Another possible extension of the formalism is provided by Coulomb
systems in an arbitrary $\nu$-dimensional Euclidean space with 
the Coulomb potential $1/r^{\nu-2}$ which are of mathematical interest
\cite{Sari76,Rougerie13}.

\begin{acknowledgments}
L. \v{S}. is grateful to LPTMS for its hospitality. 
The support received from the Grant VEGA No. 2/0015/15 is acknowledged. 
\end{acknowledgments}

\appendix*

\section{Small-$\eta$ expansion of $z(\eta)$} \label{Appendix}
For all three geometries $\alpha=0,1,2$, we consider the PB equation
\begin{equation}
\frac{{\rm d}^2 \phi(r)}{{\rm d}r^2} + 
\frac{\alpha}{r} \frac{{\rm d}\phi(r)}{{\rm d}r} 
= \frac{z}{R^2} {\rm e}^{\phi(r)}
\end{equation} 
with the BCs $\lim_{r\to 0} r^{\alpha}\phi'(r) = 0$ and $R\phi'(R)=\eta$.
The gauge is fixed to $\phi(R)=0$.
We assume that the electric potential $\phi(r)$ is small.

In the leading DH order, we substitute the exponential $\exp[\phi(r)]$
by unity:
\begin{equation}
\frac{{\rm d}^2 \phi(r)}{{\rm d}r^2} + 
\frac{\alpha}{r} \frac{{\rm d}\phi(r)}{{\rm d}r} 
= \frac{z}{R^2} .
\end{equation} 
The general solution of this differential equation reads
\begin{equation}
\phi(r) = \frac{z}{2 R^2 (1+\alpha)} \frac{r^2}{R^2} +
c_1 \frac{r^{1-\alpha}}{1-\alpha} + c_2 .
\end{equation}
The BC $\lim_{r\to 0} r^{\alpha}\phi'(r) = 0$ implies $c_1=0$.
The gauge $\phi(R)=0$ sets $c_2=-z/[2R^2(1+\alpha)]$. 
The BC $R\phi'(R)=\eta$ leads to
\begin{equation}
z = (1+\alpha) \eta .
\end{equation}

The next order corresponds to $\exp[\phi(r)]\sim 1 + \phi(r)$.
The resulting equation
\begin{equation}
\frac{{\rm d}^2 \phi(r)}{{\rm d}r^2} + 
\frac{\alpha}{r} \frac{{\rm d}\phi(r)}{{\rm d}r} 
= \frac{z}{R^2} \left[ 1 + \phi(r) \right]
\end{equation} 
has the solution
\begin{equation}
\phi(r) = - 1 + c r^{\frac{1-\alpha}{2}} 
J_{\frac{\alpha-1}{2}}\left(-{\rm i}\frac{\sqrt{z}}{R} r\right)
\end{equation}
with the Bessel function of the first kind $J$,
which automatically fulfills the BC $\lim_{r\to 0} r^{\alpha}\phi'(r) = 0$.
The gauge fixes $c=R^{\frac{\alpha-1}{2}}/J_{\frac{\alpha-1}{2}}(-{\rm i}\sqrt{z})$.
Since
\begin{eqnarray}
R \phi'(R) & = & {\rm i}\sqrt{z} \frac{J_{\frac{\alpha+1}{2}}(-{\rm i}\sqrt{z})}{
J_{\frac{\alpha-1}{2}}(-{\rm i}\sqrt{z})} \nonumber \\ & = &
\frac{z}{1+\alpha} - \frac{z^2}{(1+\alpha)^2(3+\alpha)} + {\cal O}(z^3) ,
\end{eqnarray}
the BC $R\phi'(R)=\eta$ leads to the expansion 
\begin{equation} \label{fexp}
z = (1+\alpha) \eta + \frac{1+\alpha}{3+\alpha} \eta^2 + {\cal O}(\eta^3) .
\end{equation}

The derivation of next expansion terms is a more difficult task.

\end{document}